\documentstyle[aps,prd,floats,graphicx,twocolumn,eqsecnum]{revtex}
\def\be{\begin{equation}}
\def\ee{\end{equation}}
\def\bea{\begin{eqnarray}}
\def\eea{\end{eqnarray}}

\def\Om{\Omega}
\newcommand{\gsim}{\mbox{\raisebox{-1.ex}{$\stackrel
     {\textstyle>}{\textstyle\sim}$}}}
\newcommand{\lsim}{\mbox{\raisebox{-1.ex}{$\stackrel
     {\textstyle<}{\textstyle \sim}$}}}
\newcommand{\square}{\kern1pt\vbox{\hrule height  1.2pt\hbox{\vrule
width 1.2pt\hskip 3pt \vbox{\vskip 6pt}\hskip  3pt\vrule width
0.6pt}\hrule height 0.6pt}\kern1pt}
\begin{document}
\draft
 \twocolumn[\hsize\textwidth\columnwidth\hsize\csname
 @twocolumnfalse\endcsname

\preprint{}

\title{Condensate cosmology -- dark energy from dark matter}

\author{Bruce A.~Bassett$^1$, Martin Kunz$^{2,3}$, David Parkinson$^1$ and 
Carlo Ungarelli$^{1,4}$}

\address{${}^1$Institute of Cosmology and Gravitation, University of 
Portsmouth, Portsmouth~PO1~2EG, UK\\
${}^2$Astrophysics, Department of Physics, 1 Keble Road, Oxford University,
Oxford OX1 3RH, UK \\
${}^3$Astronomy Centre, CPES, University of Sussex, Brighton, BN1 9QJ, UK\\
${}^4$School of Physics and Astronomy, University of Birmingham
Edgbaston, Birmingham, B15 2TT, UK}

\date{\today}
\maketitle

%
%

\begin{abstract}
        Imagine a scenario in which the dark energy forms via the
        condensation of dark matter at some low redshift. The Compton
        wavelength therefore changes from small to very large at
        the transition, unlike quintessence or metamorphosis. 
        We study CMB, large scale structure, supernova and radio
        galaxy constraints on condensation by performing a 4 parameter
        likelihood analysis over the Hubble constant and the three
        parameters associated
        with $Q$, the condensate field: $\Omega_{Q}$, $w_f$ and $z_{t}$
        (energy density and equation of state today, and redshift of
        transition). Condensation roughly interpolates 
        between $\Lambda$CDM (for large $z_t$) and sCDM (low $z_t$) and 
        provides a slightly better fit to the data than
        $\Lambda$CDM. We confirm that
        there is no degeneracy in the CMB between $H$ and $z_t$ and  
        discuss the implications 
        of late-time transitions for the Lyman-$\alpha$ forest. Finally
        we discuss the nonlinear phase of both condensation and 
        metamorphosis, which is much more interesting than in standard
        quintessence models.       
\end{abstract}
\vskip 1pc \pacs{98.80.Es, 04.62.+v, 98.80.Cq \ 
\hfill \ \ astro-ph/0211303 \ \ }
\vskip 1.6pc 
 ]

\section{Introduction}

The observational evidence that the universe is accelerating today
\cite{Ber,AGR} changes the study of inflation from cosmic archaeology
to real-time ghost hunting and implies that our universe is dominated
by energy with negative pressure 
or that Einstein's General Relativity is incorrect on large scales.  

In this paper we discard the second possibility, but consider 
the idea that the dark energy comes from the condensation of dark 
matter (which we take to be CDM). We show that the current data does not 
particularly favour the simplest model of condensation over the 
standard dark energy candidate: quintessence\cite{CDS}.
Nevertheless it has a very interesting nonlinear picture of the universe, 
as we describe in section V, and may serve as an archetype for other 
scenarios for dark energy. 

On a more general note, our analysis is perhaps the first detailed 
comparison with  current data of a dark  energy model with a 
Compton wavelength, $\lambda_c \equiv (V'')^{-1/2}$,  and speed of 
sound, $c_s^2 = \delta p/\delta \rho$, 
which change radically with time. By contrast, quintessence
and metamorphosis \cite{PR} both involve scalar fields that are light at 
all times, though k-essence \cite{kes} by way of contrast, has 
a speed of sound which is low at decoupling, which may be 
observable in the CMB \cite{sound}. General issues with regard to the 
measurement of the dark energy speed of sound were discussed in \cite{HETW}.

The idea that acceleration might be induced by a condensate  
has been discussed by several authors. It has been suggested that 
a scalar condensate might arise in an effective field theory description of 
gravity \cite{BV}. Condensates have also been 
invoked to explain inflation in earlier work\cite{BM,Parker93}. 
These condensates would naturally form at very high energies, but from 
condensed matter theory we are also familiar with condensates forming at 
very low temperatures. 

An example is a super-conductive material where, below a threshold
temperature, a phase transition takes place and the electrons (which
are fermions) form Cooper pairs. Those Cooper pairs  are collectively 
described by a boson field. In our cosmological scenario, when the
temperature of the Universe reaches a suitable value, the cold dark
matter, which could be associated to a fermion (e.g. the neutralino),
might undergo a similar phase transition where a condensate of
fermion-fermion pairs would emerge.  Such pairs would then be
described by an effective scalar field theory.  Initially, the cold dark 
matter (CDM) scales as dust ($w_{0}=0$).  If after the transition the final 
equation of state of the condensate $w_{f}<-1/3$, then the scalar 
field dynamics will drive the universe into an accelerating phase. 

The condensation mechanism has also been advocated recently in order
to explain the late acceleration of the universe. In particular,
in~\cite{CaldiC} it has been observed that a cosmological constant
with the suitable energy scale ($\Lambda \sim (10^{-3} \mbox{eV})^4$)
can be explained by neutrino condensation. Furthermore, in~\cite{QCD} it 
has been found that in supersymmetric, non-Abelian gauge theories 
with chiral/anti-chiral matter fields the condensation of those 
chiral fields in the low-temperature/strong-coupling regime yields 
a scalar field which has an inverse power-law potential and 
therefore is a possible quintessence field candidate. Similar ideas are 
explored in \cite{gold}. 

\begin{figure}[h]
\begin{center}
\includegraphics[width=70mm]{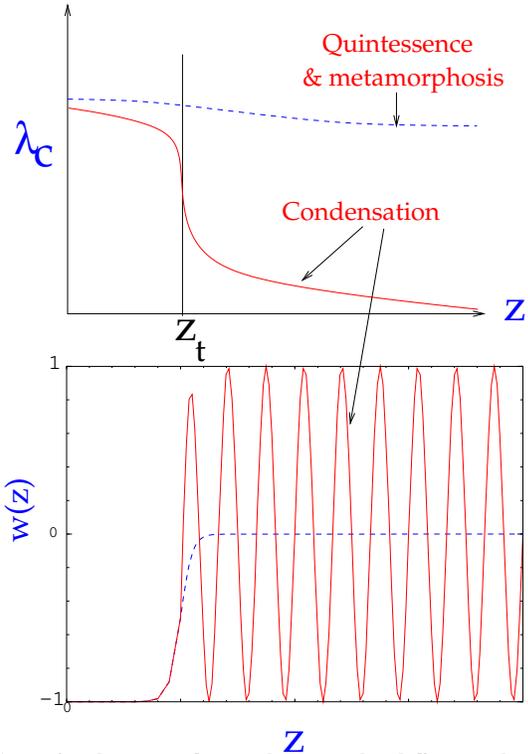} \\
\caption[eos]{\label{eos} 
A schematic figure showing the difference between condensation and
metamorphosis \& quintessence. {\bf Top:} the Compton wavelength
$\lambda_c \equiv (V'')^{-1/2}$. {\bf Bottom:} The equation of state
$w(z)$. In quintessence and metamorphosis $w(z)$ has some slowly
varying value during radiation and matter domination. In condensation,
$w(z)$ only vanishes on averaging over many oscillations, as for
standard dark matter. After the transition at $z = z_t$ condensation
becomes very similar to both metamorphosis and quintessence. For this
example we chose $w_f = -1$.} 
\end{center}
\end{figure}

In an earlier paper we studied vacuum metamorphosis which also involves
a late-time transition. We found that CMB, LSS and SN-Ia data prefer a 
transition around $z \sim 1.5-2$\cite{meta}.  One of the aims of this
paper is to check whether the data simply prefer a late-time
transition in the background cosmology (which controls the onset of
acceleration) or are also fundamentally sensitive to the nature of
the perturbations. To this end we compare metamorphosis and condensation. 

Within our phenomenological descriptions both of these have identical 
background dynamics. However, the perturbation physics is
fundamentally different. In metamorphosis the scalar field is very
light ($m_Q \sim 10^{-33} eV$) at all times\footnote{In this paper 
we will denote the dark energy field by $Q$, 
whether it be quintessence, metamorphosis or condensation. Since the context
should always be clear this should not lead to confusion.}, 
both before and after the transition at $z_t$ where $w(z)$ changes. In
condensation by contrast, the mass of the field is large before
the transition ($z > z_t$). We choose $m >  1 GeV$ for $z > z_t$ 
as appropriate for CDM, 
with a compton wavelength corresponding to nuclear scales,
$\lambda_C \lsim 1 {\rm fm}$.
Other choices, such as warm dark matter 
($m \sim 1 keV$) are certainly possible. After the
transition ($z < z_t$) the scalar field becomes very light, with a Compton
wavelength larger than $1000 {\rm Mpc}$. We will show that the data are
rather sensitive to these resulting differences in perturbation dynamics.

There are at least four reasons why it is worth analysing 
a condensation mechanism for the origin of quintessence: 
Firstly, it can provide a link between dark energy and dark matter through 
a known physical phenomenon. This minimalist philosophy has been recently
exploited in trying to obtain acceleration and clustering from the same 
source field \cite{paddy} with a scale-dependent equation of state. It is also 
the basic driving idea behind the use of the Chaplygin gas which has equation
of state $p \propto -\rho^{-\alpha}$, $\alpha > 0$ and which interpolates 
between dark matter and dark energy \cite{chap0,chap1,chapneven}, quite similar
to the class of models which we study here. But it appears 
to have problems with age estimates of high-$z$ objects \cite{chap2}.
 
Even more severe is that a recent analysis of fluctuations in 
Chaplygin-gas models finds that all models which differ
significantly from effective $\Lambda$CDM are ruled out by the matter power
spectrum \cite{sandvik} or the CMB \cite{bean}. Sandvik {\em et al} 
find that if the sound velocity of the dark matter is not zero, 
its fluctuations are either unstable towards collapse ($c_s^2<0$) or
start to oscillate ($c_s^2>0$).
Although the equation of state of the Chaplygin gas and our condensation
model are quite similar, 
it must be pointed out, that the condensation model is immune to 
this problem. The Compton wavelength of the dark energy becomes very
large just at the transition (when the sound speed starts to differ from
zero), and the small-wavelength fluctuations in this component are suppressed 
and cannot grow out of control.

Secondly, condensation of CDM into a 
scalar field may give rise to scaling solutions
characterised by a fixed fraction of the condensate relative to the CDM
independent of redshift.
Additionally the characteristic temperature of a condensation process 
is usually low, so such mechanisms naturally explains why a quintessence field 
would dominate only at low redshifts, although a condensation occurring 
in the future cannot be ruled out. In general a condensation mechanism for 
late-time acceleration would provide a partial solution to the coincidence 
problem. 

Thirdly, it is an example of a model where the effective mass of
the dark energy field changes strongly. It allows us to study the
influence of such a mechanism on CMB and LSS data, which probe the
fluctuations as well as the background geometry.

Finally, one interesting feature of condensation is its nonlinear
origin. Hence, CDM condensation may give rise to significant
modifications for the {\em background cosmology} in regions of high
density (such as clusters of galaxies) as compared to
voids. Therefore, there exists the possibility that condensation may be 
relevant to the core-cusp problem of CDM \cite{SS}, an idea recently explored
in \cite{ALS}.

The aim of this paper is to analyse in detail how the condensation  
mechanism can be tested/falsified through its effects on CMB anisotropy, 
the matter power spectrum and the supernova
luminosity distance. We also compare this class of models with standard 
and non-standard quintessence models such as vacuum metamorphosis 
\cite{PR}. The outline of the paper is the 
following: in the next section we describe the class of phenomenological 
models which we have adopted for the condensation and how condensation of cold 
dark matter into a scalar field affects CMB anisotropies and the matter power 
spectrum. In section III we test the model predictions with CMB, large 
scale structure and supernova data.  In section IV we discuss how 
condensation differs from conventional quintessence and metamorphosis.
We go on to examine how further constraints could be made with large 
scale structure analysis, angular-diameter distance to radio sources 
measurements and detection of fluctuations in the dark energy.  In 
section V we discuss the non-linear evolutionary phase of the 
cosmologies.  Section VI contains our conclusions and discussion. 

\section{The phenomenological models}
\label{stwo}

In this paper we will compare and contrast three phenomenological models, 
particularly focusing on the first two:

\begin{itemize}
\item {\em Metamorphosis}\cite{PR} --  a light scalar field undergoes 
a transition in its equation of state from {\em exactly pressure free} 
dust to $w_f < -0.3$ at some  redshift $z_t$. This is appropriate for 
approximately describing many quintessence models \cite{CC2}.

\item {\em Condensation}  --  A heavy scalar field with quadratic potential 
(e.g. CDM, {\em dust on average}) undergoes a transition in $w(z)$ to 
$w_f < -0.3$ and a very light mass $\sim 10^{-33} eV$ at $z_t$. Before the 
transition there are no light scalar fields or cosmological constant. 

\item {\em Evaporation} --  The inverse of condensation. At high redshifts
$z > z_t$, there is no CDM. There is only a light scalar field and baryons. 
At $z_t$ the scalar field condensate evaporates to yield the CDM, while 
$w(z)$ becomes negative to drive acceleration at late times. 
\end{itemize}

We will now discuss in more detail these three cosmic scenarios. 

\subsection{Metamorphosis}

The archetype of metamorphosis is the model proposed by Parker and Raval
\cite{PR} in which non-perturbative quantum effects drive a transition in 
the equation of state of the scalar field. Metamorphosis was studied 
in detail in \cite{meta} where it was found that the data 
prefers a late-time transition. 
The equation of state which we use to parametrise both metamorphosis and 
condensation is:
\be
w(z) = w_{0} + \frac{(w_{f}-w_{0})}{1+\exp((z-z_{t})/\Delta)}
\label{wz}
\ee
where we assume that $w_{0}=0$ and $\Delta=30 z_t$ where $\Delta$ controls
the width of the transition. This equation of state is then fed into
a modified Boltzmann code which uses the reconstructed scalar field potential
along the background trajectory of the field, Q(t) (see \cite{meta}).
For an analysis of the effects of $\Delta$ on
the CMB see \cite{CBUC}.  For condensation this equation holds exactly for $z < z_t$ 
but only on average for $z > z_t$.

In terms of the scalar field dynamics, this class of 
models is characterised by three free parameters: the redshift $z_t$ at 
which the transition takes place, the fraction $\Omega_{Q}$  
of the total energy density today due to the scalar field and the final 
value $w_{f}$ of the equation of state of the scalar field. 

We note that in fact this metamorphosis describes rather accurately a
wide range of standard quintessence models\cite{CC2},
particularly the Albrecht-Skordis model\cite{AS}. 
\footnote{The parametrisation of $w(z)$ in eq. (\ref{wz}) 
is interesting in terms of the statefinder pair $\{r,s\}$
introduced in \cite{statef}. For a rapid transition, $\{r,s\} =
\{1,0\}$ for $z>z_t$ and $\sim \{1,0\}$ for $z < z_t$ if $w_f \sim
-1$. The pair differ strongly from these values only near $z = z_t$
where $\dot{w}(z)$ is large. }

The main contrast between metamorphosis and condensation is that in the 
metamorphosis case the Compton wavelength of the field $Q$ is very large
at all times, despite having $w = 0$ at early times.  In condensation, as 
mentioned earlier, before the 
transition  $\langle w \rangle = 0$ {\bf only}, the Compton
wavelength is very short until the transition, after which the field
is very similar to the metamorphosis case; see fig. (\ref{eos}).

\subsection{Condensation} 

The distinctive feature of condensate models is that the dark energy 
emerges from a condensation process which occurs 
when the universe reaches a certain critical temperature. 
This implies that above the
redshift $z_t$ at which the condensation occurs 
the dynamics of the background is determined by photons, baryons, CDM, 
and neutrinos, while below the redshift $z_t$  
cosmological dynamics is also determined by the dark energy field $Q$, 
its energy density $\rho_Q$ and its equation of state $w(z)$. 

In general the details of the condensation will depend on the precise model
used. We are not even guaranteed that such a condensate process is possible
for realistic beyond-the-standard-model physics.
Hence for simplicity and 
due to our ignorance of the detailed microphysics we 
shall assume that the condensation occurs instantaneously on a
constant energy hypersurface. Furthermore we  
impose that the total energy density is conserved  through the transition. 
Since the transition is instantaneous this implies that the total CDM-$Q$
density perturbation is conserved. 

\bea
&&\rho^{-}_{CDM} = \rho^{+}_{CDM} + \rho^{+}_{Q} 
\label{eqcont} \\
&&\delta\rho^{-}_{CDM} = \delta\rho^{+}_{Q} + \delta\rho^{+}_{CDM}
\label{eqcontp}
\eea
The fraction $F$ of CDM energy density which is transferred to the scalar 
field $Q$ ($\rho^{+}_{Q}=F\rho^{-}_{CDM}$) can be easily inferred 
from Eq.~(\ref{eqcont}) 
\be
F= \frac{\Omega_{Q}}{\Omega_{CDM}}\Psi_w(z_t)
\frac{1}{1+\frac{\Omega_{Q}^{o}}{\Omega^{o}_{CDM}}\Psi_w(z_t)}
\label{fraction}
\ee
 where $\Omega_{Q},\Omega_{CDM}$ are respectively the energy densities 
of the scalar field and CDM in units of the critical energy density today.
The function $\Psi_w(z_t)$ is given by 
\be
\Psi_w(z_t)=e^{3\int^{z_t}_0\frac{dz}{(1+z)}w_Q(z)} .
\label{psieq}
\ee
In the following we shall consider a particular profile for the equation of 
state of the scalar field which coincides with that used for metamorphosis, 
i.e. equation (\ref{wz}). 

Using Eq.~(\ref{fraction},\ref{psieq}) it is straightforward to see how 
dark energy domination is addressed within condensation. To achieve
$\Omega_{Q}/\Omega_{CDM} = O(1)$ implies that 
\be
F\sim e^{3\int^{z_t}_0\frac{dz}{(1+z)}w_Q(z)} .
\ee
If we assume that $w(z) < 0$ for $z < z_t$, condensation at high redshift has 
to produce a small fraction of scalar field energy density. Similarly, the 
more negative the equation of state of the scalar field is, the less 
efficient the condensation process has to be. 

In order to analyse the imprint of the condensation process on the CMB 
anisotropies and on the matter power spectra we need to solve the evolution 
equation for the scalar field fluctuations $\delta Q$. 
The simplest initial conditions for 
the perturbations of the scalar field $Q$ are 
$\delta Q = \delta\dot{Q} = 0$. However, since we are considering an 
instantaneous transition, this choice of initial conditions may not preserve 
adiabaticity \footnote{We assume the primordial spectrum of fluctuations is
purely adiabatic.}. 
In particular a discontinuity in the time-derivatives of the scalar field 
$Q$ could generate a non-zero value of the intrinsic 
entropy perturbation of the scalar field, $S_{Q\dot{Q}}$: 
\be
S_{Q\dot{Q}} \equiv \frac{\delta Q}{\dot{Q}} - \frac{\delta\dot{Q}}{\ddot{Q}}\,.
\ee
This always vanishes on large scales \cite{preh} 
but could alter sub-Hubble dynamics
if not treated carefully. 

In order to fix the initial conditions for the scalar field perturbations, 
we follow the method developed in \cite{DandM} and we match the 
fluctuations of the scalar field $Q$ along a constant energy 
density hypersurface, which appears the most appropriate for a 
condensation process. In particular, the matching conditions for the 
perturbations are obtained by requiring the continuity of the induced 
metric and the extrinsic curvature. In the synchronous gauge and in a 
frame comoving with the CDM fluid, this implies that, for each Fourier mode 
$\delta Q_k$ the quantity $\theta_Q$ defined as 
\be
(\rho_Q (1+ w_Q))\theta_Q \equiv -k^2\dot{Q}\delta Q_k
\label{theta}
\ee
must be continuous at the transition. Using the 
conditions~(\ref{eqcont},\ref{eqcontp}) we have 
$\theta_{Q}=\delta Q = 0$ across the transition and  $\delta\dot{Q}$ 
can be determined in terms of 
$\dot{Q}$ and $\rho^{-}_{CDM}$ by requiring $S_{Q\dot{Q}} = 0$. 
From then on, the evolution of the perturbations is calculated
using the same modified Boltzmann code approach as for the metamorphosis
model\cite{meta}.

\subsection{Evaporation}\label{evapor}

Since we have considered the possibility that dark matter condenses at some
low temperature into dark energy, it seems natural to consider the inverse 
process.  This would be the evaporation of a condensate to yield all 
or part of the dark matter of the universe.
Less severe forms of evaporation have been studied within the context 
of Affleck-Dine baryogenesis \cite{Moroi} where a scalar condensate 
decays to give the required baryon asymmetry. 
 

In its extreme form (no dark matter before evaporation) 
and if the  evaporation occurs suddenly and completely at $z_t$ then 
a clear disadvantage of evaporation is that the universe does 
not accelerate today since $\Omega_Q = 0$.  
It therefore makes sense to study evaporation models in 
which the transition has not completed by today and has a large width, 
$\Delta$ or not all the scalar field evaporates into dark matter. 
Can we put limits on such models? 
Let us consider the simplest model in which the there is no dark matter 
before the transition redshift $z_t$. For $w_f < -0.5$ and for 
$z \gg z_t$ the universe is essentially a baryon-dominated flat universe 
since the condensate energy density is negligible at high redshifts \footnote{
Though note that if $w = 0$ and $\Omega_Q$ is not negligible at decoupling this 
will not be true.}.

This places large constraints on the model since it is extremely difficult 
to reconcile CMB data, big bang nucleosynthesis (BBN), a flat universe and 
SN-Ia data (the incompatibility is $3\sigma$ or more \cite{GMS} in a $\Lambda$
baryon dominated model). In addition, the large baryon content of the universe tends to cause 
very prominent oscillations in the matter power spectrum that are not
strongly observed in current data. This latter test is less 
straightforward to apply to evaporation since the nascent 
dark matter might tend to smooth out the oscillations if it is
warm enough to have a sufficiently large free-streaming length.
 
A very interesting point is that evaporation would lead to 
a strongly time-dependent bias around the transition redshift, 
an effect which naively links to the work of Magliochetti {\em et al}
\cite{manuela} who found a strong redshift-dependence of the bias. 

From the results of Griffiths {\em et al} \cite{GMS} we can conservatively 
infer that, for $w_f \sim -1$, a low $z_t$ is unacceptable for evaporation 
unless $w_0$ is close to $0$.  In fact, since the 
secondary acoustic peaks are sensitive to the 
baryon and dark matter content, it is likely that best fit to the CMB 
would occur for $z_t > 1000$. 

Such values would be disfavoured from SN-Ia observations unless the 
transition were extremely wide so that there still remained a significant 
component  with negative equation of state at low redshifts. 
As such, only "glacier evaporation" models would seem to be observationally
viable. However, the evaporation process of the condensate in such a slow 
manner is physically realistic and evaporation has some interesting 
nonlinear implications which we discuss in section \ref{nonlinear}. 
We will not, however, explicitly compare evaporation models with current data,
leaving that to future work. 

\section{Comparison with observations} \label{sec:like}

\subsection{Data and analysis method}
In order to constrain/falsify our phenomenological models of condensation, 
we compare its predictions with a number of observations. In particular we 
consider three different kinds of datasets: CMB anisotropy, 
large scale structure (LSS) and type-1a supernova (SN-Ia) data. 
For the CMB anisotropies, we 
use the decorrelated COBE-DMR data \cite{dmr} and the recent data from 
the DASI \cite{DASI}, MAXIMA \cite{maxima}, BOOMERanG \cite{Nett} and CBI \cite{Pearson} 
experiments. The total number of data points is $n_{\rm CMB}=58$, 
with $l$ ranging from $2$ to $1235$  for the first four sets, and the mosaic 
data from CBI extends this to $1900$. In the analysis, we take into account 
calibration uncertainties for DASI, MAXIMA, BOOMERanG and CBI. 
\begin{figure}[h]
\begin{center}
\label{alldata}
\includegraphics[width=70mm]{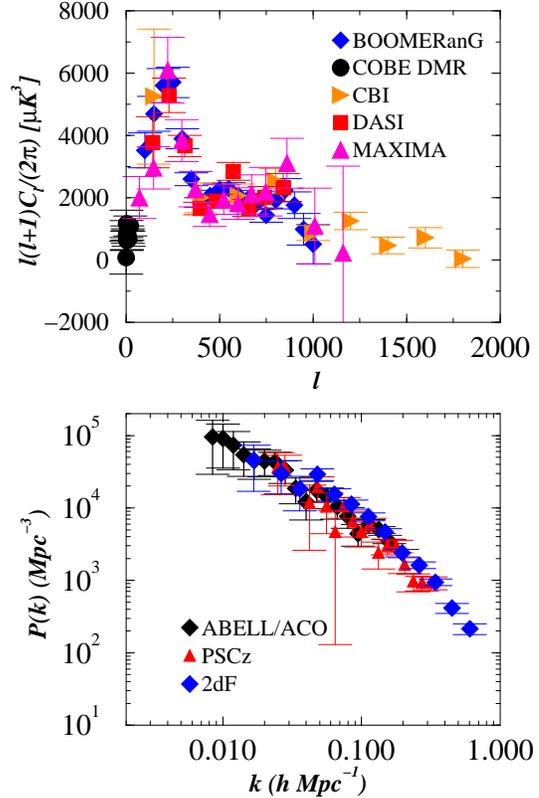} \\
\caption[alldata]{A compilation of our data for  
 the CMB (top) and LSS (bottom) used in our analysis. The PSCz and 2df
 are galaxy surveys while the Abell/ACO points are for cluster data. The 
 supernovae data are as described in \cite{meta}.
 } 
\end{center}
\end{figure}

As far as LSS 
data are concerned, we use the matter power spectrum estimated from the 
following surveys: 2dF 100k redshift survey \cite{2df}, IRAS PSCz 0.6
Jy \cite{pscz}, and Abell/ACO cluster survey \cite{Abell}. In order to
avoid possible effects due to non-linear contaminations, we fix a
cut-off $k_{\rm max}=0.2h/$Mpc. The total number of data points is
$n_{\rm LSS}=48$. Finally, we use the redshift-binned SN-Ia data
\cite{sn1a} from HZT and SCP ($n_{\rm SN}=7$). Additionally we also
studied the angular-diameter distance to radio sources
\cite{Gurvits99}. This method is potentially able to probe a large
range of redshifts, current data already covers $0.1 < z < 4$.  All of
our models as well as $\Lambda$CDM are perfectly consistent with
current data, which provides a valuable test. Unfortunately we will
have to wait for larger samples before we can constrain our additional
parameters significantly.

We performed a likelihood analysis by varying the three parameters directly 
related to the scalar field ($\Omega_Q,w_f,z_t$) and the Hubble parameter, 
$H_0 \equiv 100 h_0\,{\rm km},{\rm s}^{-1},{\rm Mpc}^{-1}$. For the
other parameters 
we chose the following values: $\omega_b \equiv \Omega_b h^2 =0.02$, 
$n_s=1$, $n_t=0$, $\tau=0$, $\Omega_{tot}=1$, where $\tau$ is the reionisation 
optical depth and $n_{s,t}$ are the spectral indices for the scalar and 
tensor perturbations respectively. We also include the 
``standard'' effective number number of neutrinos (3.04) and we fix the 
cold dark matter density as $\Omega_{\rm CDM}=1-\omega_b/(h_0^2)-\Omega_Q$. 
In our analysis we set the amplitude of the tensor perturbations to zero. 

In order to compute the likelihood functions for the various parameters we 
follow the usual procedure. We compute the power spectra $C_\ell$ of the CMB 
anisotropies and the CDM matter power spectra $P(k)$ for each set of parameters 
over the four-dimensional space $(\Omega_Q,z_t,w_f,H_0)$. We then evaluate 
the $\chi^2$ at each grid point. The CMB anisotropy spectra and the matter 
power spectra are related through their respective amplitude normalisations. 
We found that the CMB data fixes quite well the overall amplitude for each 
model. As far as the LSS data are concerned we allowed a bias $1/5 <b < 5$ 
for 2dF and PSCz and $1/9<b<9$ for Abell/ACO since clusters 
are expected to be more biased than galaxies. We choose these
large bias values since the biasing mechanism is still not very well
understood especially in the context of non-standard dark matter models, 
and since (as argued in section \ref{nonlinear}) the
possibility of strongly nonlinear physics enters on cluster scales.

We then finally determine the 
one-dimensional and two-dimensional likelihoods by integrating over the 
other parameters. 
\subsection{Results}
\begin{figure}[h]
\begin{center}
\includegraphics[width=80mm]{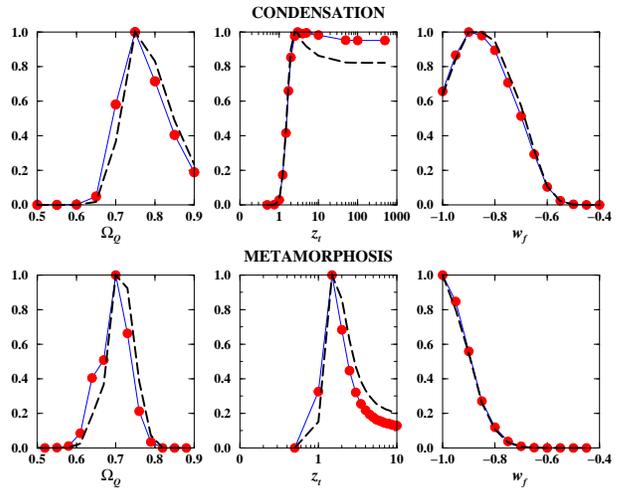} \\
\caption[1dplots]{\label{1dplotsall} 
The 1d-likelihood curves ( 
{\bf Top}: Condensation, {\bf bottom}:  Metamorphosis) for the 
scalar field parameters $(\Omega_Q,z_t,w_f)$ obtained combining all the 
CMB, LSS and  SN-Ia data. 
The dashed curves are the likelihoods obtained by introducing a 
Gaussian prior on $H_0$ ($h_0=0.72\pm 0.08$).}
\end{center}
\end{figure}

\begin{figure}
\begin{center}
\includegraphics[width=80mm]{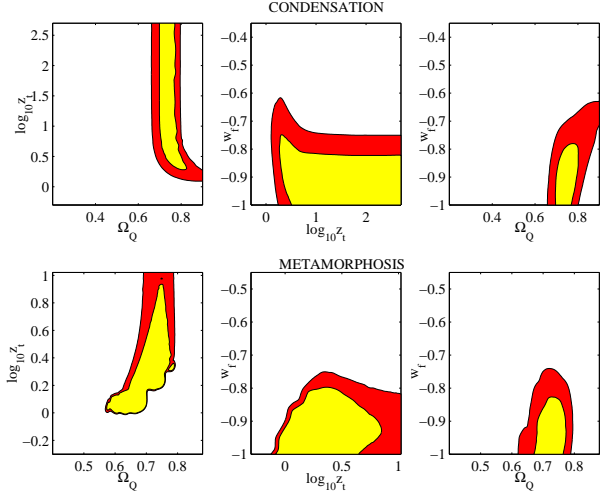} \\
\caption[2dplots]{\label{2dplotcmall} 
The 2d-likelihood  contours 
({\bf Top}: Condensation, {\bf bottom}:  Metamorphosis) for the 
scalar field parameters $(\Omega_Q,z_t,w_f)$ obtained combining all the 
CMB, LSS and  SN-Ia data.  }
\end{center}
\end{figure}

Our main results are shown in figures (\ref{1dplotsall}) and (\ref{2dplotcmall}).
Both models prefer a energy density $\Omega_Q$ close to $0.7$ ($0.7$
for metamorphosis, $0.75$ for condensation).  Both models also prefer
a final value of the equation of state $w_{f} < -0.8$. A big
difference is seen in the likelihood curve for $z_t$.  Whereas
metamorphosis favours $z_t \sim 1.5 - 2.5$, in condensation a late
transition ($z_t < 1.5$) is strongly disfavoured.

Overall the condensate model and the metamorphosis models both have 
slightly better best-fit $\chi^2$ than the best $\Lambda$CDM model, although
this is to be expected since we have extra parameters and is not 
highly significant.  The
condensation best fit $\chi^2 = 79.3$, with parameters $z_t=4$, $w_f =
-0.95$ and $\Omega_Q = 0.75$, while the $\Lambda$CDM best fit $\chi^2=
84.9$ (for $\Omega_{\Lambda}=0.73$).  The best fit for metamorphosis has
$\chi^2= 78.8$, with parameters $z_t=1.5$, $w_f=-1$ and $\Omega_Q=0.73$.

The following
sections discuss in detail the CMB and LSS results. It should
be noted that the supernova data only probes the background geometry, which
is the same for both metamorphosis and condensation. Since it
is also independent of the Hubble constant, the results
of \cite{meta} are unchanged, except that we have 
updated the brightness of the supernovae at $z \sim 1.7$ to account for 
lensing \cite{lenssn}. Namely, there is a slight preference
for a low-$z$ transition, but the relatively large error bars and lack of
points at $z > 1$  do not allow strong 
conclusions. Section \ref{sec:dist} discusses more
general aspects of using distance measurements to probe the background
geometry of the universe.


\section{Comparison of condensation and metamorphosis}
\begin{figure}
\begin{center}
\includegraphics[width=80mm]{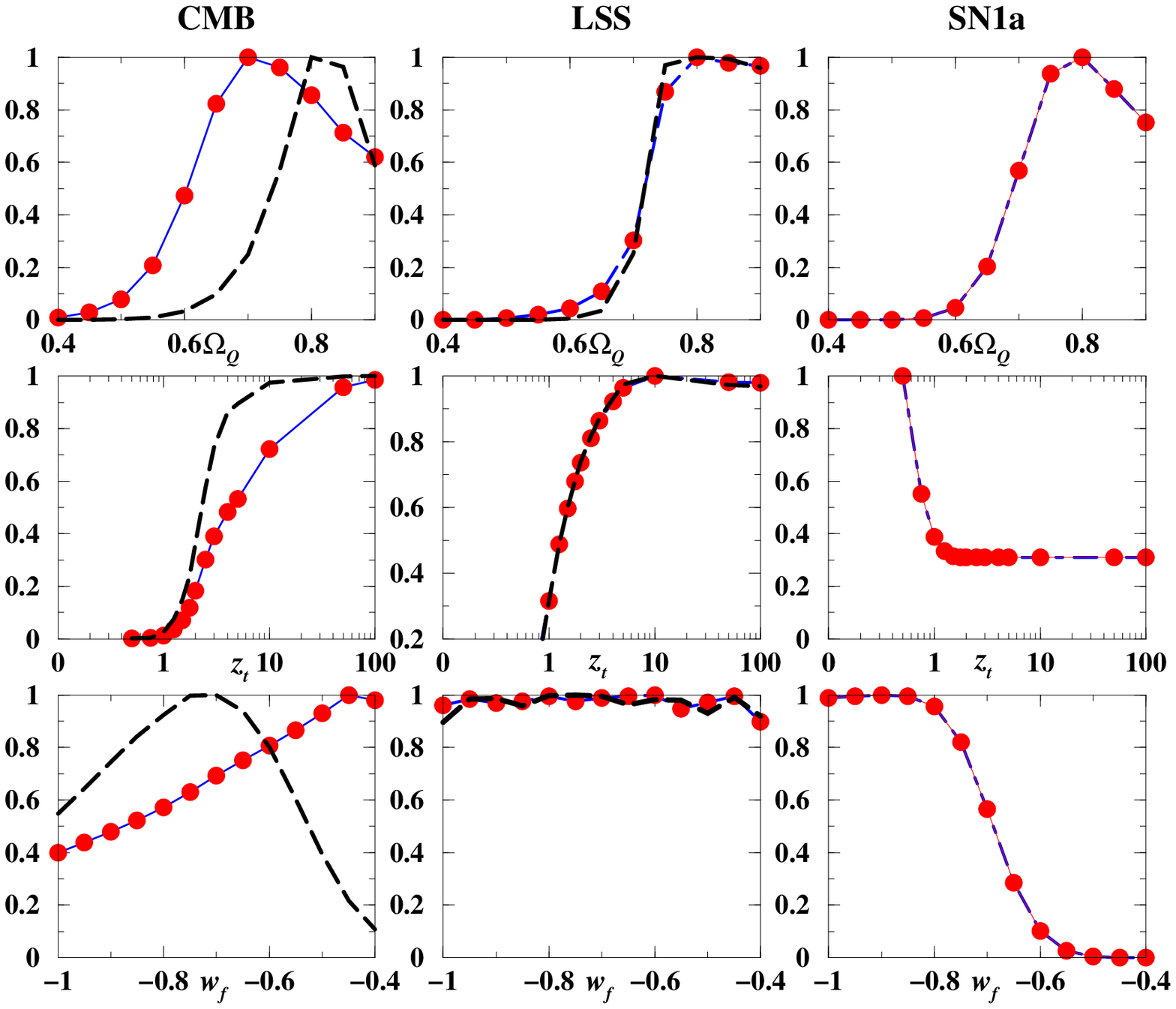} \\
\caption[1dplots]{\label{1dplotsc} {\bf Condensation:}  
the 1d-likelihood functions for the 
scalar field parameters $(\Omega_Q,z_t,w_f)$. 
Left-column: CMB. Middle column: LSS. Right column: SN-Ia.
 The thick solid orange curves are the likelihoods obtained by 
 introducing a Gaussian prior on $H_0$
($h_0=0.72\pm 0.08$).}
\end{center}
\begin{center}
\includegraphics[width=80mm]{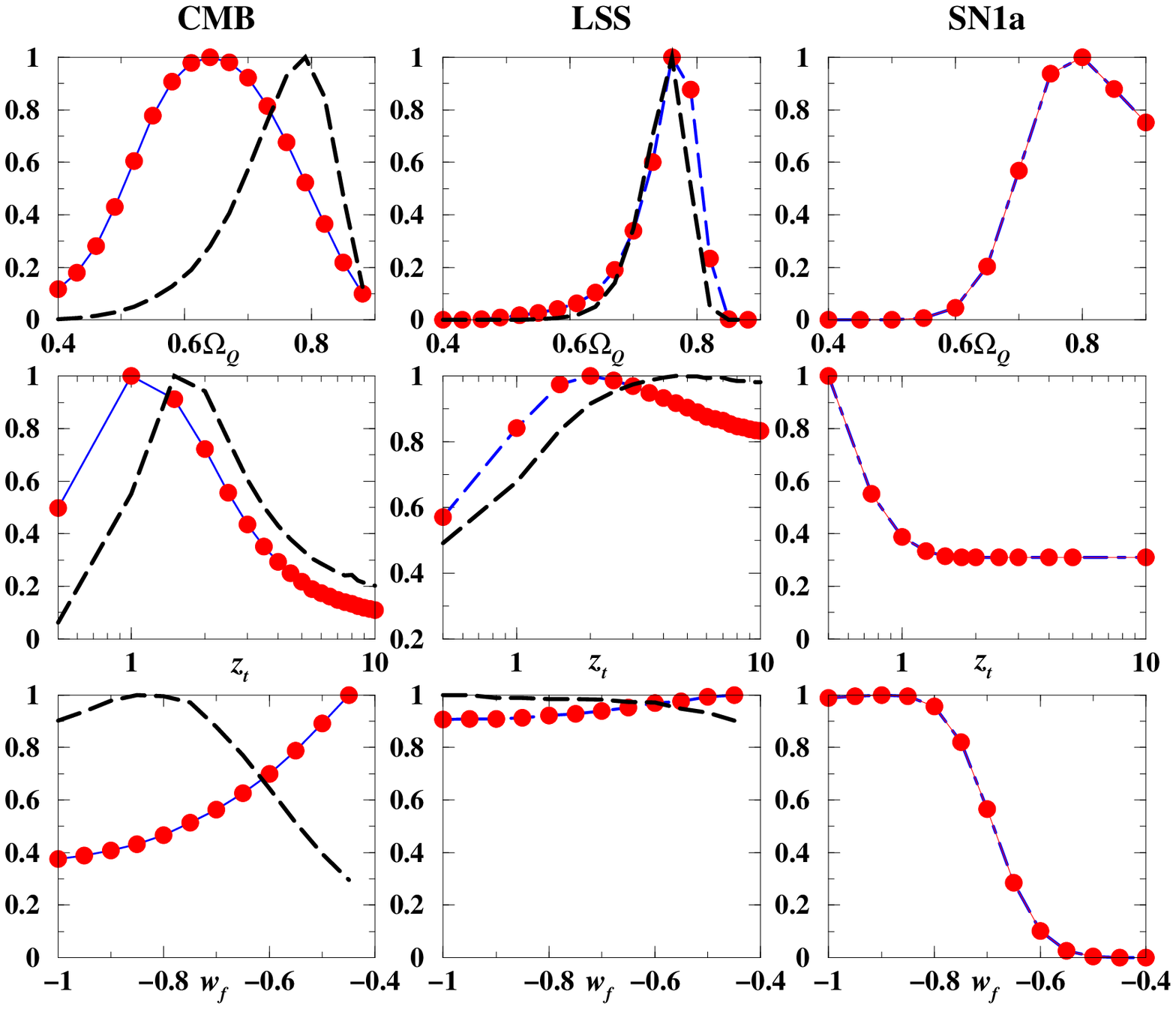} \\
\caption[1dplots]{\label{1dplotsm} {\bf Metamorphosis:}  
the 1d-likelihood functions for the 
scalar field parameters $(\Omega_Q,z_t,w_f)$. 
Left-column: CMB. Middle column: LSS. Right column: SN-Ia. The thick 
solid orange curves are the likelihoods obtained by introducing a 
Gaussian prior on $H_0$
($h_0=0.72\pm 0.08$).}
\end{center}
\end{figure}

\subsection{CMB}
The likelihood plots for condensation and metamorphosis are very
similar when considering  $\Omega_{Q}$ and
$w_{f}$.  The major difference comes with the redshift of the transition
$z_t$.  Although the background dynamics are identical, Metamorphosis 
favours a transition of $z_t \simeq 2$, whereas condensation strongly
disfavours $z_t < 1.5$.  

\begin{figure}[h]
\begin{center}
\includegraphics[width=70mm]{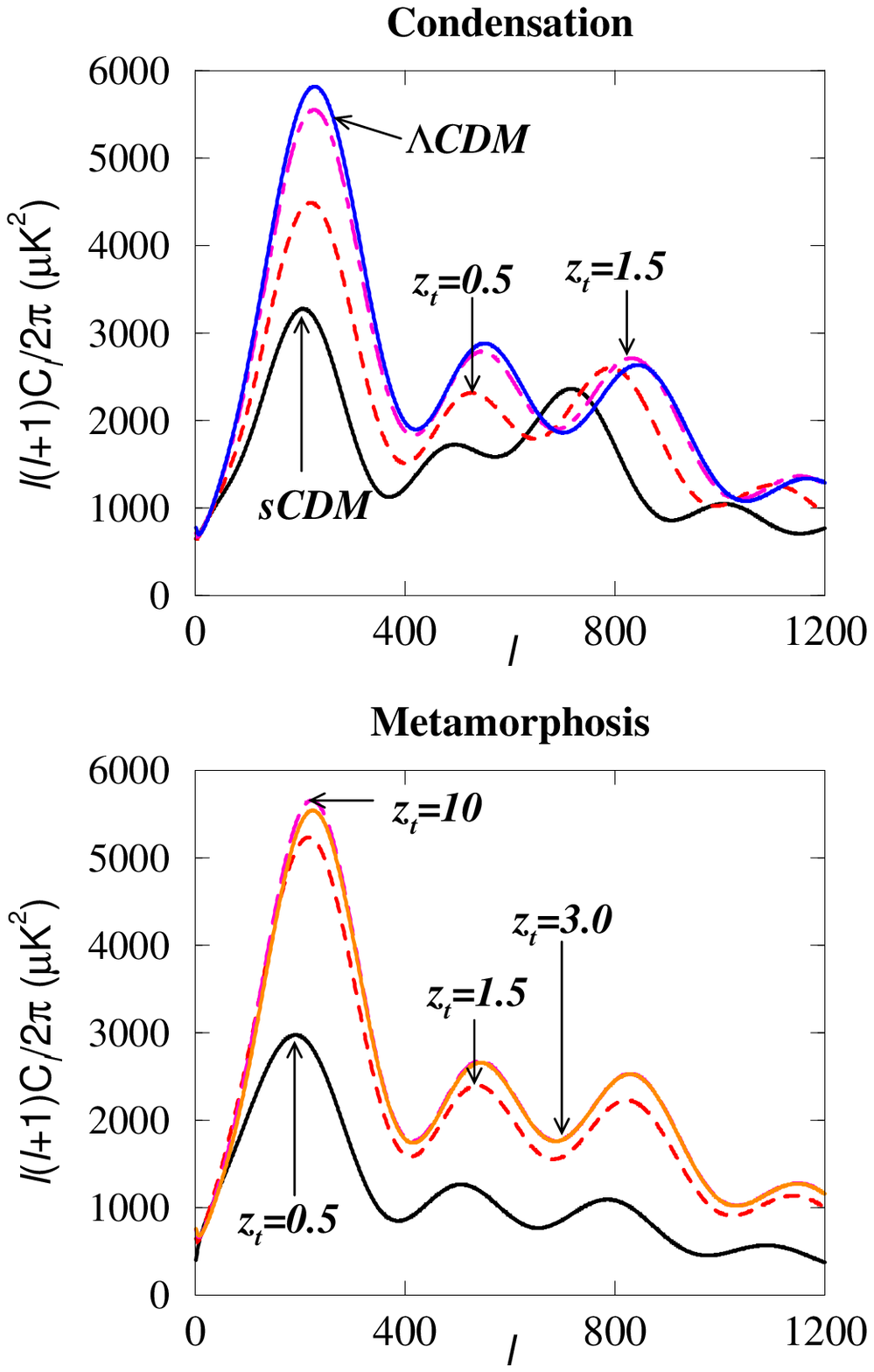} \\
\caption[cl_zt]{\label{clat} The $C_{\ell}$ spectra for condensation (top)
  and metamorphosis (bottom) as a function of $z_t$.  For comparison in the
  condensation case sCDM $(\Omega = 1, \Lambda = 0)$  
  is the lowest curve at $\ell = 200$,
  $\Lambda$CDM the highest. The other parameters are fixed to be
  $\Omega_Q = \Omega_{\Lambda} = 0.7$, $h_0 = 0.63$ and $w_f = -0.95$.
}
\end{center}
\end{figure}

To explain this we need to take a closer look at the respective
CMB angular power spectra, see fig.~\ref{clat}.
Metamorphosis, like quintessence, has an
almost trivial effect on the CMB relative to $\Lambda$CDM, 
simply through the ISW effect
which changes the Sachs-Wolfe plateau and hence the relative
height of the peaks through the CMB normalisation on large scales.
The overall form of the CMB power spectrum of metamorphosis is
not changed.

The effect of $z_t$ on the acoustic peaks of
condensation is more subtle.
Although the total density of all species is
unchanged, the phase-transition means that the change in the CDM
density is not continuous.  Changing $z_t$ changes the CDM
density at decoupling and the redshift of matter-radiation equality, 
even though its value, $\Omega_{CDM} \simeq 1 - \Omega_Q$, is unchanged today.  This
alters the gravitational potential wells, although the effect is only
significant for small $z_t$, where the amount of CDM that is
condensing out is at a maximum. 
Decreasing the amount
of CDM changes the time of matter-radiation equality, which boosts
the amplitude of all oscillations through an ISW effect. The third
peak is moved to slightly higher $\ell$ and thereby damped more,
making it less high than the second peak
\cite{huS,huD,durrer}.
Overall, the condensation
CMB power spectrum effectively interpolates between the two extreme
cases of sCDM  ($\Lambda = 0, \Omega = 1$) and $\Lambda$CDM.

\begin{figure}[h]
\begin{center}
\includegraphics[width=70mm]{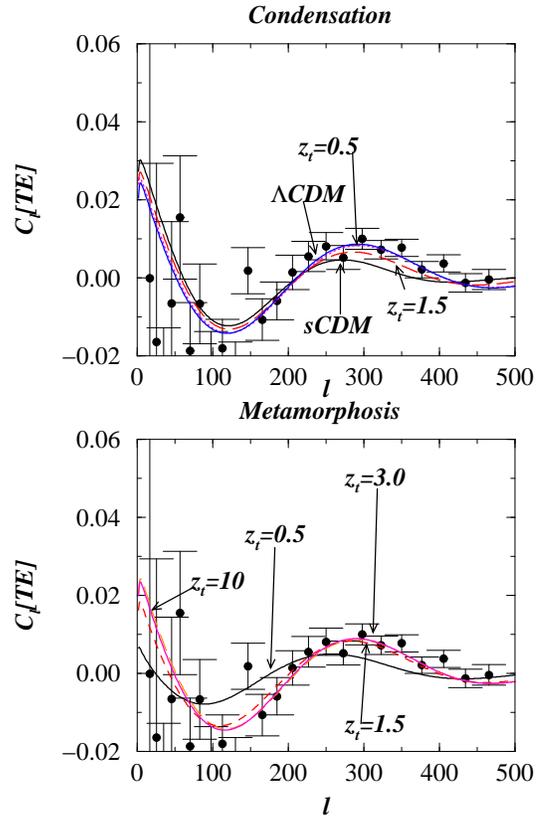} \\
\caption[cl_zt]{\label{TE} The $C^{TE}_{\ell}$ temperature-polarisation 
spectra for condensation (top)
  and metamorphosis (bottom) as a function of $z_t$ vs the WMAP data.  
  For comparison in the
  condensation case sCDM $(\Omega = 1, \Lambda = 0)$  and 
  $\Lambda$CDM curves are included. 
}
\end{center}
\end{figure}

For interest's sake we also plot in fig (\ref{TE}) the predictions for the
temperature-polarisation ($TE$) cross-power spectra for condensation and 
metamorphosis and compare them with the data from the first year of WMAP 
data \cite{wmap}. Currently the data cannot distinguish between the models, 
especially as the curves are sensitive to reionisation history which will
in turn depend on $z_t$ etc...  

\subsection{LSS}

The large scale structure data also strongly disfavours any value of
the transition redshift $z_t < 2$ for condensation.  In contrast, 
the case of metamorphosis, $z_t$ seems to be almost independent of the 
LSS data, as shown in figs. (\ref{1dplotsc},\ref{1dplotsm}). Why is this?

\begin{figure}[h]
\begin{center}
\label{tk}
\includegraphics[width=80mm]{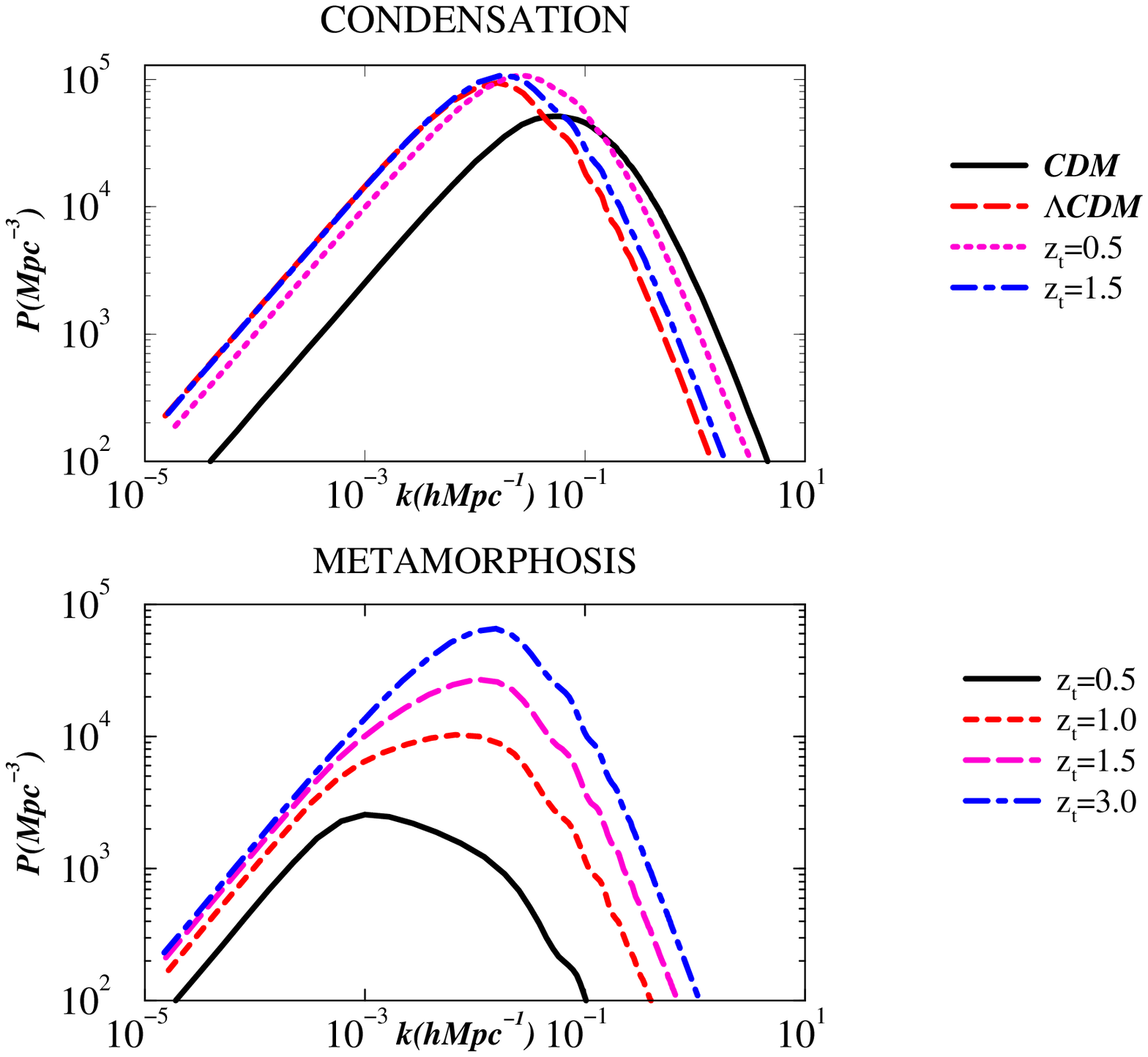} \\
\caption[transfer]{\label{transfer} The COBE normalised matter 
  power spectra for
  condensation (top) and metamorphosis (bottom) as a function of
  $z_t$, with sCDM and $\Lambda$CDM to compare with in the condensation
  case.The other parameters are fixed to be
  $\Omega_Q = \Omega_{\Lambda} = 0.7$, $h_0 = 0.63$ and $w_f = -0.95$.
}
\end{center}
\end{figure}

Comparing the transfer functions, $T(k)$, for the metamorphosis and condensation 
models with $\Lambda$CDM and sCDM (where the universe is flat and CDM 
dominated without cosmological constant so $\Omega_{\rm CDM} = 1 - \Omega_b$) is 
informative.  When normalised to give the correct COBE normalisation, 
sCDM gives a slightly smaller amount of power on
large scales, but gives more power on small scales.

In condensation the Compton wavelength is
small before the transition, and only becomes big after the
transition. Clustering therefore does take place on all scales 
for $z > z_t$. For $z< z_t$, the growth of clustering on small scales 
ceases due to the large $\lambda_c$ but this also occurs on all scales 
since the universe starts to accelerate.  The closer that 
transition is to today, the more clustering on small scales that can occur, 
resulting in a larger mass dispersion, $\sigma_8$. Again, the power spectrum
interpolates between sCDM $(z_t \rightarrow 0)$ and $\Lambda$CDM ($z_t \gg 1$).
Metamorphosis, in
contrast, is a light scalar field
before and after the transition,  with little change in the Compton
wavelength and so the large scale structure is only weakly dependent on
the transition redshift, as discussed in \cite{meta}.

But why does the LSS data disfavour a low $z_t$ for condensation? We have not
computed $\sigma_8$ for all these models (we do so below for 
specific parameter values), so the reason cannot be that. Similarly, our 
bias upper
limits of $5$ and $9$ for galaxy and clusters respectively are unlikely to 
have a real impact in excluding low-$z_t$ models. 

The key lies in the change to
the redshift of the matter-radiation equality, $z_{eq}$, which changes
significantly with $z_t$ for $z_t < 5$ and provides the dominant effect on
the shape of the power spectrum. Comparing the LSS data (fig. \ref{alldata})
and the condensation $T(k)$ (fig. \ref{tk}) we see that the sCDM power spectrum
turns over around $k \sim 0.1 h Mpc^{-1}$ while for $z_t = 0.5$ the turnover
starts around $0.05 h Mpc^{-1}$. In comparison, the data do not show any definite 
sign of a turnover, even out to $k = 0.01 h Mpc^{-1}$. Hence both 
the CMB and the LSS disfavour condensation with $z_t < 1.5$. 
The same conclusion is likely if one does include $\sigma_8$ constraints.

The overall normalisations of the power spectra are not trivial, 
since they are linked to
the COBE normalisation of the CMB power spectrum. We can read
it off the high-$k$ region of fig.~\ref{transfer}. While 
condensation lies between
$\Lambda$CDM and sCDM as expected, we have to remember that the
ISW effect of metamorphosis changes the normalisation strongly
and leads to a low value of $\sigma_8$ \cite{CBUC}.

\begin{figure}[h]
\begin{center}
\includegraphics[width=80mm]{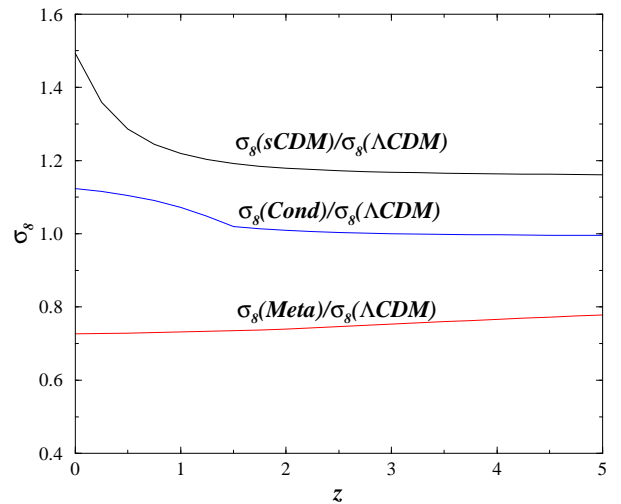} \\
\caption[sigma8]{\label{sigma8} The ratio of $\sigma_8$ (mass dispersion 
smoothed over a radius
of $8h^{-1}Mpc$) as a function of redshift for the different models, as compared
to $\Lambda$CDM. $\Omega_\Lambda=0.7$ for the 
$\Lambda$CDM model, and $0$ for sCDM. $\Omega_Q=0.7$, 
$w_{f}=-0.95$ and $z_t=1.5$ for the condensation and 
metamorphosis models. The effect of the growth of small-scale perturbations 
due to the small Compton wavelength in the condensation case, for $z > z_t$ is clear.} 
\end{center}
\end{figure}

The differences in perturbation evolution mentioned earlier can 
also be seen in fig.~\ref{sigma8}, which shows $\sigma_8$
(relative to the one of a $\Lambda$CDM universe) as a function of 
redshift for $z_t$ = 1.5 and $w_f = -0.95$. 
Standard CDM starts to produce more structure,
especially at low redshifts where a cosmological constant would
start dominating. Condensation behaves for $z>z_t$ like sCDM,
but its behaviour changes  at $z_t$ 
roughly when the universe begins to accelerate, thereby forcing the
linear perturbations to stop growing. 

The redshift dependence of $\sigma_8$ and the bias factor can therefore be 
used to distinguish these models.  Magliochetti{\em et al} 
\cite{manuela} found a strong signal of a redshift-dependent $\sigma_8$ 
for $z > 2$ and redshift dependent bias.  The redshift dependent bias $b(z)$ 
is expected in our models since the baryons will take some time to 
respond to the sudden change in the dark matter characteristics. We leave 
this interesting issue to future work. 

As a final point we note the strong differences between the CMB 1d likelihood curves 
in figs. (\ref{1dplotsc}) and (\ref{1dplotsm}) with and without the HST prior on 
the Hubble constant. In particular, $w_f$ changes drastically reflecting the well-known
CMB degeneracy between $H$ and $w_f$.

\subsection{Distance measurements} \label{sec:dist}

One important tool to constrain the equation is state is the determination
of distances as a function of redshift, e.g. with supernovae. The basic
building block is the comoving distance (for a flat universe)
\be
d_C(z) \equiv c(\eta_0 - \eta(z)) = \frac{c}{a_0} \int_0^z \frac{dz}{H(z)}
\ee
where $\eta(z)$ is the conformal time and
\be
H(z) = H_0 \left[ \sum_i \Om_i (1+z)^{3(1+w_i)} \right]^{1/2} .
\ee
$i$ indexes the constituents of the energy density in 
the universe, eg. radiation ($w=1/3$), matter ($w=0$),
curvature ($w=-1/3$) and quintessence type contributions, where
$w$ can vary as a function of $z$ like in our models.
\begin{figure}[h]
\begin{center}
\includegraphics[width=40mm]{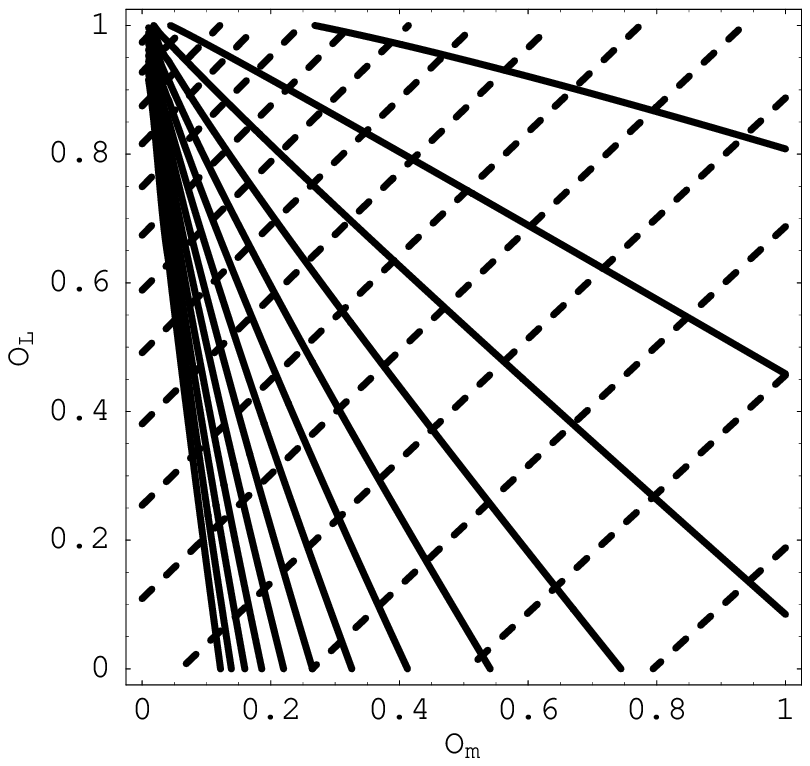}
\includegraphics[width=40mm]{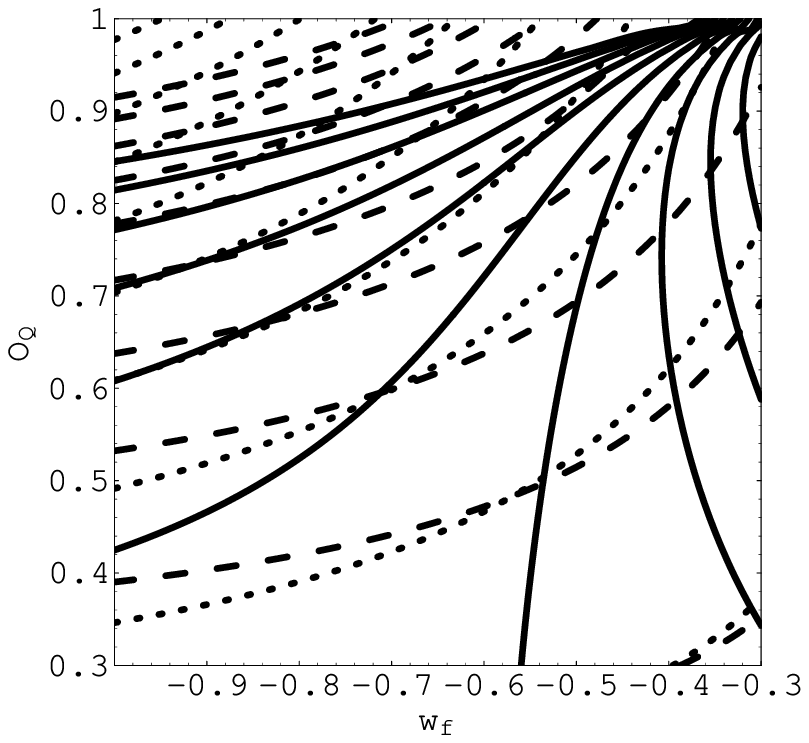}  \\
\caption[deg1]{\label{deg1} The left figure shows the degeneracies
for measuring $\Omega_m$ versus $\Omega_\Lambda$ with supernovae
at $z=1$ (dashed lines) and the position of the CMB peaks (solid
lines). The degeneracies are clearly orthogonal and an accurate
measurement of both variables is possible. The right figure shows
the same for $w$ and $\Omega_Q$ and measurements at $z=1$ (dotted),
$3$ (dashed) as well as the CMB peak position (solid). 
The likelihood is very flat in the $w$ direction, making it
difficult to pinpoint it precisely, especially for a very negative
equation of state. The models shown here have a constant equation
of state ($z_t$ is very large).}
\end{center}
\end{figure}
The apparent distance to a bright object such as a 
supernova is then determined by the luminosity
distance $d_L = (1+z) d_C$, while the distance to an object of a
given physical length (like radio sources, see \cite{Gurvits99}) is described by 
the angular diameter distance $d_A = d_C / (1+z)$ \footnote{$d_L$ and $d_A$ are, of
course, equivalent and related via the reciprocity relation.}. It is also
possible to use the CMB in a similar way: the angle subtended by
the sound horizon at recombination gives us a fixed scale, and
we can use it to determine the geometry between $\eta_{\mathrm{rec}}$
and today. Since the horizon size is a comoving quantity and not
a physical quantity, it is $d_C(z_{\mathrm{rec}})$ which enters
in this case. (See e.g. \cite{doranphd} and references therein 
for a detailed discussion.)

While it seems at a first glance that the distance measures depend
on the value of the Hubble constant $H_0$, this is not so: all data
available compares two lengths and thereby divides out any dependence
on $H_0$. In the CMB case, the size of the sound horizon at last
scattering depends linearly on $1/H_0$ just like $d_C$; in measuring
the luminosity distance to SN-Ia, the normalisation of the distance
ladder is treated as a nuisance parameter and effectively determined
in units of $1/h$, as is the case for the physical reference length
when working with the angular size of radio sources.

\begin{figure}[h]
\begin{center}
\includegraphics[width=40mm]{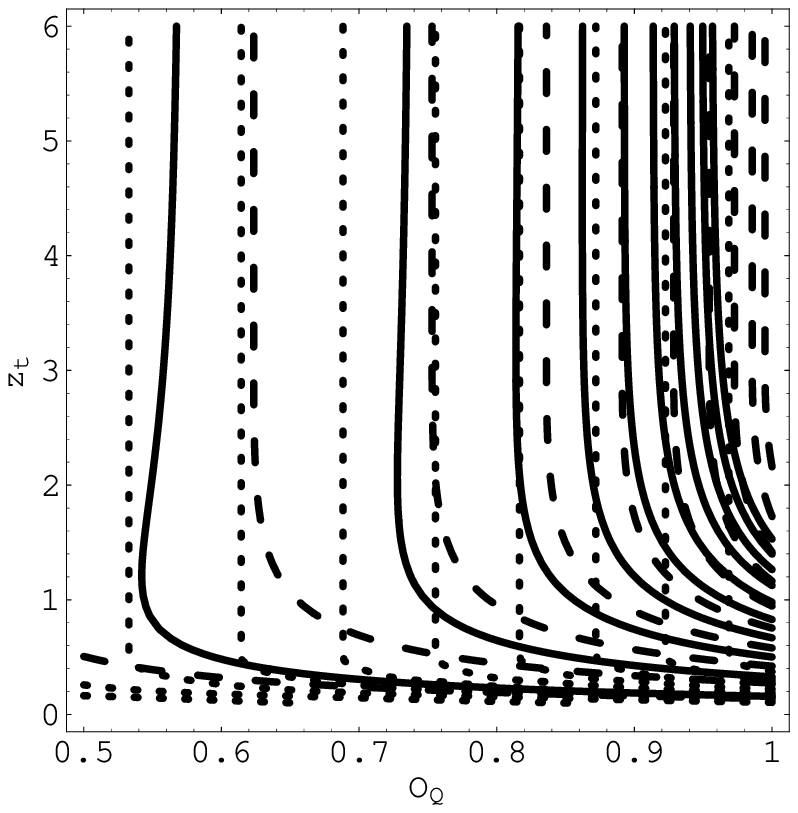}
\includegraphics[width=40mm]{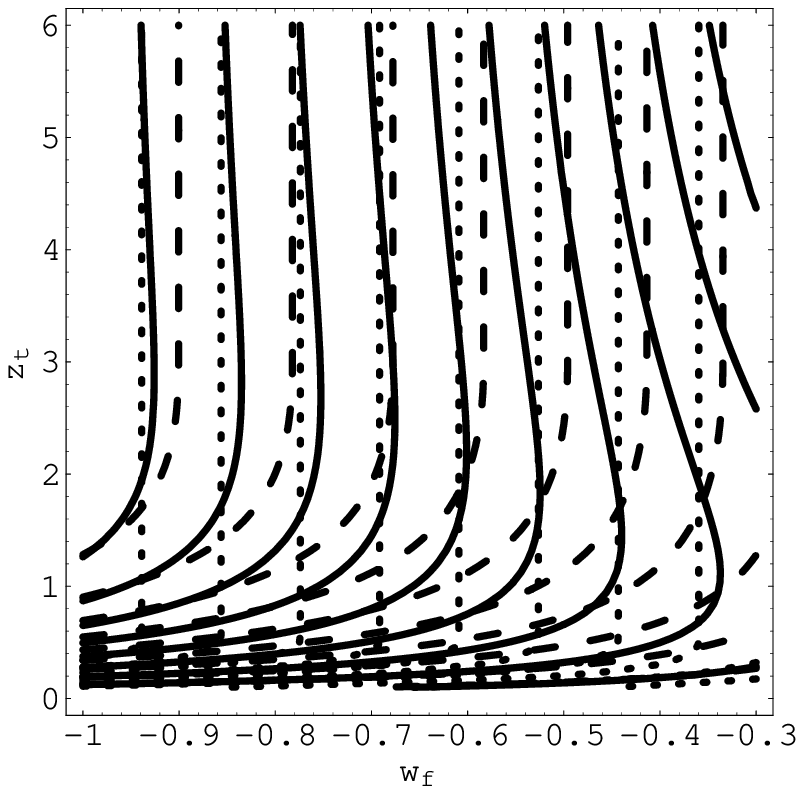}  \\
\caption[deg2]{\label{deg2} The left figure shows the degeneracies
for a model with $w_f = -1$ and observations at $z=0.5$ (dotted), 
$z=3$ (dashed) and the CMB (solid).
The right figure shows the 
same for $\Omega_Q = 0.5$.
A priori it is necessary to bracket the phase transition. As the
right plot shows, high-$z$ distance data can help finding the
transition in certain cases, but does usually not improve much
on the CMB data (which is effectively at $z\approx1100$).
Also, transitions
at redshifts higher than 5 will be very hard to detect unambiguously
using distance measurements.}
\end{center}
\end{figure}

It is not possible to determine several quantities from
only one data point (e.g. the location of the CMB peaks)\footnote{
Of course the full CMB power spectrum contains many more observables,
thereby breaking some of the degeneracies.}. Measuring
distances over a range of $z$ values as done for supernovae and
radio galaxies, improves the situation somewhat, as does combining
different data sets. Indeed, a low $z_t$ can be strongly constrained
by data on both sides of the transition (see figure \ref{deg2}). 
A high $z_t \gsim 5$ may be impossible to detect given the
current limits on $w$, since the dark energy becomes irrelevant
very quickly.

Figure \ref{deg1} also reinforces what the full likelihood
figures in section \ref{sec:like} have already shown: $\Omega_Q$ can
be determined to a much better accuracy than $w$, since distance
measurements depend only quite weakly on its value.

It must be emphasised at this point that especially the CMB
degeneracies depend strongly on the quantities which are being
kept constant. If we fix $\Om_m h^2$ instead of the Hubble constant
(and therefore vary $h$ as a function of $\Om_m$), the curves
look substantially different (see e.g. \cite{huterer}). Nonetheless, the
overall conclusions remain the same.

\subsection{Change in the comoving number density of the Lyman-$\alpha$ absorbers}

The number of Lyman $\alpha$ absorption lines per unit redshift,
$dn/dz$, has been of interest to observers for a long time. It is
sensitive both to the background evolution and to details of the linear 
and nonlinear density perturbation evolution\cite{TLSE}. This makes
it a possible probe of quintessence type models. Unfortunately, the
non-gravitational effects make this instrument very difficult to
analyse and interpret reliably and would require high-resolution hydrodynamic
N-body simulations. Nevertheless we illustrate the underlying point with a 
simple model. 

Observers use a comoving volume element with a fixed physical radius $r_p$ 
and comoving length $dr_c$ \cite{sargent}. As $r_c = c \eta$, we find that
$dr_c/dz=1/H(z)$, and since $r_c = (1+z) r_p$ we find that
\be
dV = dr_c^3 \propto (1+z)^2/H(z) dr_p^2 dz.
\ee

\begin{figure}[h]
\begin{center}
\includegraphics[width=70mm]{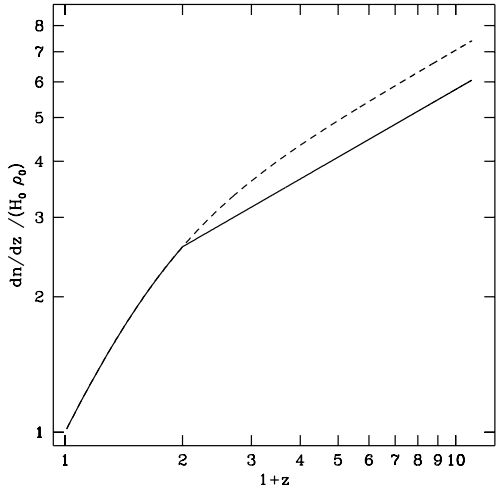}  \\
\caption[lya]{\label{fig:lya} An illustration of the break in
the exponent $\gamma$ derived from (\ref{dndz}). While a
$\Lambda$CDM universe shows a smooth transition from a steep slope
at low redshift to a slope of $\gamma = 1/2$ when CDM starts
to dominate (dashed line), a
condensation or metamorphosis model exhibits a break at the
transition redshift (here $z_t = 1$, solid line). Although Lyman $\alpha$
data is strongly influenced by non-gravitational evolution
effects, we would expect to see some kind of jump in
$d\gamma/dz$ if a transition takes place at a low enough
redshift.}
\end{center}
\end{figure}
If we naively assume that the comoving number density of Lyman
$\alpha$ absorbers is constant, so that they directly probe the
volume as a function of redshift, we set $dn(z) = \rho_0 dV(z)$
and obtain
\be
\frac{dn}{dz} = \rho_0 H_0 \frac{(1+z)^2}{H(z)} . \label{dndz}
\ee
Observers use the parametrisation $dn/dz \propto (1+z)^\gamma$.
A pure CDM universe would therefore have $\gamma = 1/2$, while
a pure $\Lambda$ universe had a $\gamma = 2$. As observers
find $\gamma > 2$ even at redshifts larger than $1$ \cite{janknecht}
we have to conclude that either the average equation of state is
extremely negative at these redshifts, or that the non-gravitational
evolution effects dominate.

If it became possible to study the cosmological evolution
only, then the density of Ly-$\alpha$ absorbers would be
 an extremely useful tool since the change of physical
density depends on $H(z)$ and not its integral (as is the
case for distance measurements and the position of the
first peak in the CMB, see section \ref{sec:dist}).
Unfortunately the actual change of the number density
is strongly influenced by evolutionary effects.
At the
moment one can speculate whether it is possible to draw
conclusion from a differential approach: figure \ref{fig:lya}
shows $dn/dz$ for a $\Lambda$CDM model and for a universe
with a transition at $z_t = 1$. The second case shows a break
in the comoving number density, to a slope with a lower
$\gamma$. This {\em change} in the shape of $dn/dz$ might
be detectable even though the actual form of the curve is
strongly modified by additional physics.

As a final comment we note that there is also evidence for a change in the 
Lyman-$\alpha$ optical depth 
at a redshift $z \sim 3.2$ \cite{bern}.

\subsection{Detection of fluctuations in the dark energy}

Even though scalar field dynamics can lead to a background evolution
very close to the one of a $\Lambda$CDM cosmology, there is a basic
difference: A cosmological constant is a smooth quantity, while
the scalar field can have fluctuations \cite{meta,bacci02,CBUC}. In
the models considered here, these fluctuations can introduce 
modifications of the $C_\ell$ spectrum of the order of a percent or more, even
in the worst case of a large Compton wavelength at all times.
Especially in the condensation
scenario, it is impossible to avoid those fluctuations, since they
are present in the dark matter. The matching conditions will
transfer them into the dark energy component. An example is provided
by the two curves for metamorphosis and condensation with $z_t=0.5$ 
in fig. \ref{clat}. Both have the same effective equation of state, but
the resulting CMB power spectra are very different.

An experiment like Planck, which is basically cosmic variance
limited at low to average $\ell$ ($\sigma_\ell \approx 1/\sqrt{2\ell+1}$),
could detect (or rule out) these fluctuations for some of the
more prominent cases (like a $z_t \ll z_{\rm LSS}$ or a small
$\lambda_C$). Even though this effect will be very difficult
to disentangle from the influence of other cosmic parameters,
it would provide an important insight into the nature of the dark energy.

\section{The nonlinear evolutionary phase} \label{nonlinear}

Both the condensation and metamorphosis models have an extremely interesting 
nonlinear portrait. Since the exact time and location of the the two 
transitions is sensitive to local conditions of density and temperature, 
areas of nonlinearity can significantly delay the onset of the transition. 

Let us consider metamorphosis first. In metamorphosis the transition 
occurs when the Ricci scalar grows to be of order $-m^2/(\xi - 1/6)$ 
\cite{meta}. Now $R = -\kappa T$ where $T = \rho - 3p \sim \rho$ for CDM. 
Hence, in high-density regions which will separate from the Hubble flow and 
recollapse to form bound structures like galaxy clusters, the Ricci 
scalar is more negative than in the low-density inter-galactic medium (IGM).

This implies that the IGM will be the first place where the metamorphosis 
transition occurs. Inside rich clusters, it may even occur that the local
nonlinear growth of density perturbation overcomes the average FLRW 
growth of $R$ towards zero, and causes it locally to begin decreasing again. 
In these regions the metamorphosis transition might never take place.

The first implication of this is radical - the perturbed FLRW formalism 
for describing the universe rapidly becomes invalid as the universe now 
consists of a fluid with an inhomogeneous background equation of state -
negative in voids and positive in rich clusters. 

This will lead rapidly to a strongly scale dependent bias since
density perturbations in voids and in clusters will evolve very
differently. On scales smaller than the inhomogeneity in the equation
of state, the perturbed FLRW approximation is good. Inside clusters
perturbations will grow as they would in a flat, unaccelerating CDM
model while in voids the perturbations will feel the expansion. 

This has important implications since the 
issue of cluster abundance versus redshift 
suddenly becomes an extremely complex problem. It is no longer clear 
that clusters will be assembled at low-redshifts as happens in 
open or $\Lambda$-dominated universes due to the freeze-out in 
perturbation growth. Hence our linear calculations of $\sigma_8$ 
versus redshift 
must be carefully examined and may be wrong due to significant 
non-linear corrections. 

In this vein it is interesting to note that even $\Omega_Q \rightarrow 1$
is not strongly excluded in the condensation case. An obvious
test of this would be to examine galaxy rotation curves at low and high 
redshifts; one might expect the rotation curves at $z = 0$ to show the 
lack of dark matter. However, due to the nonlinear nature of 
the condensation, the galaxies might never undergo the condensation process
and hence might well be dark matter dominated despite most or all of the 
dark matter in the universe having already condensed. 
This nicely illustrates the potential subtleties involved in 
condensation. 
 
The first implication of these effects is the fact that metamorphosis 
and condensation (as we consider them) are likely to make the core-cusp 
problem of CDM {\em worse}.

On the other hand, a recent paper \cite{Loewe02} discussing Chandra
observations of the elliptical galaxy NGC 4636 finds no discrepancy
between the Chandra data and the existence of a central dark matter
cusp. The averaged central dark matter density inferred for this galaxy 
is about two orders of magnitude larger than for dwarf spirals and
low-surface brightness galaxies (where the core-cusp problem seems
most notable). Also intriguing, their fit significantly improved when
using a two-component dark matter model rather than just a single
component. 

It would also be
interesting to study the formation of halo substructure in condensation
type models. As this is a highly non-linear process, N-body simulations
are needed for reliable predictions.

\subsection{Nonlinear effects in evaporation}

Given that the condensation and metamorphosis models appear to make the 
core-cusp problems of standard $\Lambda$CDM worse due to the incomplete 
transition in regions of high density, it is attractive to 
consider evaporation models, as described in section (\ref{evapor}).

At the critical phase boundary the condensate, which exists early in the 
universe, starts to evaporate. 
At high temperatures, the universe consists (in the 
simplest models) solely of baryons and the condensate in addition to 
the radiation.

In these models, the high dark matter densities inside clusters would 
encourage a return to the condensate form and hence, assuming that the 
condensate had a long Compton wavelength, would suppress the formation of 
strong core density cusps which is perhaps the biggest current problem 
for $\Lambda$CDM. Meanwhile the IGM would remain in the normal, CDM, 
phase.

\section{Discussion and Conclusions}

In this paper we have  quantitatively studied a number of issues in detail: 

\begin{itemize}
\item We have considered a phenomenological model for dark energy which forms 
via condensation of dark matter, which we took to be CDM. By comparing 
with current CMB, LSS and SN-Ia data we show that we do find a better
fit to the data than from simple $\Lambda$CDM, but not significantly so. 

\item The condensation model has a Compton wavelength which changes from 
very small to very large around $z=z_t$, in contrast to quintessence or
metamorphosis where the Compton wavelength is always large. Condensation
therefore has growth of small-scale structure absent in the other 
models which impacts on the CMB and LSS. 
Nevertheless, the data slightly prefer metamorphosis, and
strongly favour metamorphosis if $z_t$ is forced to be small, $z_t < 1.5$. 

\item We have confirmed the results of \cite{meta} for metamorphosis even
when we allow the Hubble constant to vary:  current data favour 
a late-time transition in $w(z)$ at $z_t \sim 2$ (i.e. there is no 
degeneracy with $H$ as there is for $w_f$). The best-fit value
of $z_t$ for condensation is $z_t = 4$. 
\end{itemize}

We have also qualitatively discussed the nonlinear evolution of condensation 
and metamorphosis which is significantly more complex than quintessence.  

Supernova data can be used to study the evolution of the background
geometry quite independently of the dark energy model. It already
is able to set useful limits on the equation of state today. For
our class of models we find $w_f < -0.6$. The supernova data also
seems to prefer a late transition in $w$, but more data is needed
to draw strong conclusions. 

Future work should consider more realistic models for the condensate and 
examine the non-linear features of both metamorphosis and condensation 
in more detail.  Future data should improve the constraints 
on these models significantly.
Eventually, the study of the properties of the dark
energy might lead to important insights into the physics at high
energies which cannot be probed directly otherwise.

\section{Acknowledgements}
We thank Rob Lopez for part of the code used in our analysis. We thank 
C\'eline Boehm, Rob Crittenden,  Joe Silk,  James 
Taylor and David Wands for useful comments and discussions.
We thank Leonid Gurvits for the radio data.
MK acknowledges support from the Swiss National Science Foundation. BB and 
CU are supported in part by the PPARC grant PPA/G/S/2000/00115.
Part of the analysis was performed using the ICG SGI Origin.


\end{document}